\documentclass[longbibliography,twocolumn,aps,showpacs,showkeys,superscriptaddress,fleqn]{revtex4-1}
\usepackage{natbib}
\usepackage{graphicx}
\usepackage{bm}
\usepackage[table]{xcolor}
\usepackage{xcolor}
\usepackage{float}
\usepackage{amsmath}
\usepackage{mathtools}

\begin{document}

\title{Exact solution of generalized cooperative SIR dynamics}
\author{Fatemeh Zarei}
\affiliation{Sharif University of Technology, Tehran, Iran}

\author{Saman Moghimi-Araghi}
\affiliation{Sharif University of Technology, Tehran, Iran}	
	
\author{Fakhteh Ghanbarnejad}
\affiliation{ITP, Technische Universit\"at Berlin, Germany}
\affiliation{QLS, The Abdus Salam International Centre for Theoretical Physics (ICTP), Trieste, Italy}
\email{fakhteh.ghanbarnejad@gmail.com}

\begin{abstract}
In this paper,  we introduce a general framework for co-infection as cooperative SIR dynamics. We first solve analytically CGCG model \cite{chen2013outbreaks} and then the generalized model in symmetric scenarios. We calculate transition points, order parameter, i.e. total number of infected hosts. Also we show analytically there is a saddle-node bifurcation for two cooperative SIR dynamics and the transition is hybrid. Moreover, we investigate where symmetric solution is stable for initial fluctuations. Then we study asymmetric cases of parameters. The more asymmetry, for the primary and secondary infection rates of one pathogen in comparison to the other pathogen, can lead to the less infected hosts, the higher epidemic threshold and continuous transitions. Our model and results for co-infection in combination with super-infection \cite{nowak2006evolutionary} can open a road to model disease ecology.  
\end{abstract}

\maketitle

%%%%%%%%%%%%%%%%%%%%%%%%%%%%%%%%%%%%%%%%%%%%%%%%%%%%%%%%%%%%%%%%%%%%%%%%%%%%%%%%%%%%%%
\section{INTRODUCTION}
%%%%%%%%%%%%%%%%%%%%%%%%%%%%%%%%%%%%%%%%%%%%%%%%%%%%%%%%%%%%%%%%%%%%%%%%%%%%%%%%%%%%%%
Among natural disasters, infectious diseases have been one of the strongest threats against human \cite{hays2005epidemics}. Many infectious diseases including HIV, malaria, plague, Influenza and Lyme disease, have emerged, spread and evolved in a complex ecological system with multiple interacting hosts and pathogens \cite{faure2014Malarial,cuadros2011HIV,collinge2006disease,hudson2002ecology,schmidt2001biodiversity,ostfeld2010infectious}, however they mostly have been studied as single epidemiological phenomena \cite{keeling2011modeling,anderson1992infectious,hethcote2000mathematics}. Recently, some works have been studying interactions between pathogens in order to address important questions from perspective of statistical physics as well as mathematical epidemiology, such as: how cooperation or competition between pathogens affect evolution of the global spreading dynamics, the epidemic threshold and also the order of transitions from non-epidemic to epidemic regime \cite{chen2013outbreaks,chen2017fundamental,goel2018modelling,sanz2014dynamics}; or how the underlying topology of interactions between host individuals can alter the spreading dynamics \cite{cai2015avalanche,grassberger2016phase} or how their temporal or spatial correlations may favor coinfection \cite{rodriguez2017risk,rodriguez2018particle}.  

Also M. A. Nowak has addressed evolution of virulence, see chapter 11 in \cite{nowak2006evolutionary}; Considering Susceptible-Infected-Removed (SIR)-type dynamics \cite{kermack1927contribution}, he shows when pathogens, two or more strains for instance with different virulence, are in competition to occupy a host, the one with higher basic reproductive ratio will win. While in a super-infection scenario, in which one can outcome another one already exist in the host, selection does not maximize the basic reproductive ratio. It's discussed how many of the strains and how can be present at equilibrium. Nevertheless, evolution of co-infection in a general framework is not addressed. In this work we focus on single host population while several pathogens or strains, two or more, interact cooperatively and spread in the host population. The spreading dynamics are SIR-type and coupled synergistic to each other. Assuming the host population is well mixed, we treat the system in Mean-Field Approximations. Here we first find the exact solution of CGCG model \cite{chen2013outbreaks}, then we generalize the model to n-SIR dynamics and solve the system analytically. Finally we relax symmetric assumptions and study the dynamics.

%%%%%%%%%%%%%%%%%%%%%%%%%%%%%%%%%%%%%%%%%%%%%%%%%%%%%%%%%%%%%%%%%%%%%%%%%%%%%%%%%%%%%%
\section{two diseases: the Exact solution of mean field approximations}\label{analytic}
%%%%%%%%%%%%%%%%%%%%%%%%%%%%%%%%%%%%%%%%%%%%%%%%%%%%%%%%%%%%%%%%%%%%%%%%%%%%%%%%%%%%%%

\begin{figure}
\centerline{\includegraphics[scale=.35]{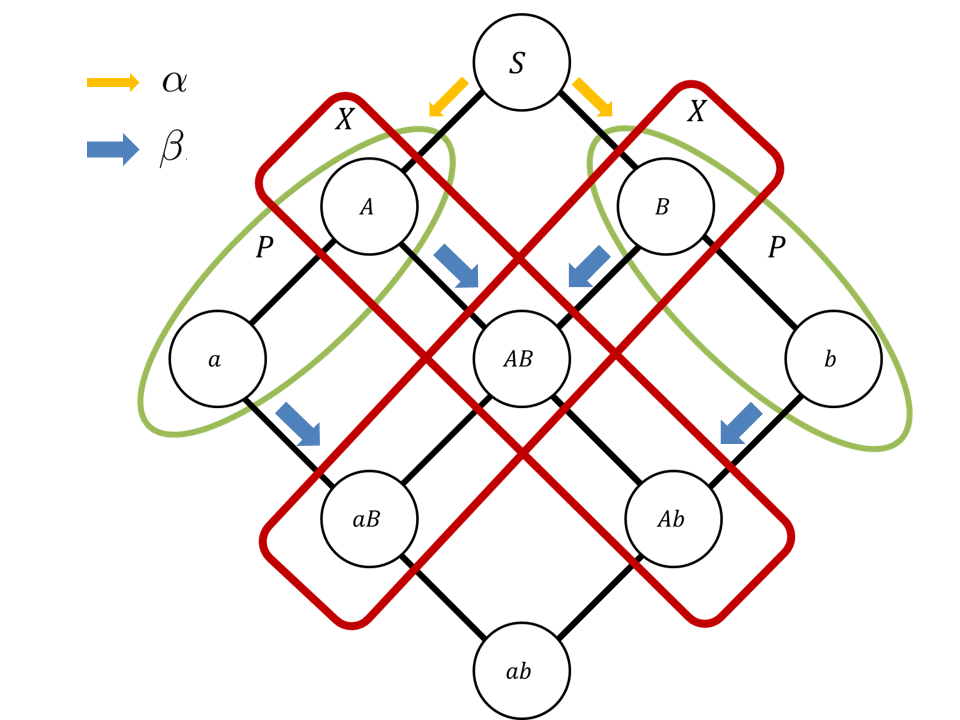}}
\caption{Flow chart of two-disease coinfection with A,B. Capital letters A and B represent infective states, lower-case letters a and b stand for “recovered” ones. X is defined as the fraction of agents that can transfer the disease A(or B) and P is defined as the fraction of agents who have experienced only one of the diseases.} 
\label{SPX}
\end{figure}

We consider the case of two diseases and call them A and B. As in \cite{chen2013outbreaks} we will denote agents who actually have the respective disease by capital letters (A, B)
, and those who had it in the past and now are immune by
lower-case letters (a, b). Assuming the two diseases have identical properties, the equations governing the dynamics would be symmetric and one can introduce the following new variables \cite{chen2013outbreaks}: $S$ the fraction of susceptible (or uninfected) agents, $X={\rm [A]+[Ab]+[AB]=[B]+[aB]+[AB]}$ the fraction of agents that can transfer the disease A(or B) and $P={\rm [A]+[a]=[B]+[b]}$ the fraction of agents who have experienced only one of the diseases (see figure \ref{SPX}). In terms of these variables the dynamical equations turn out to be:
\begin{eqnarray}\label{main-eq}
\dot{S}&=&-2 \alpha S X, \nonumber\\
 \dot{P}&=& (\alpha S-\beta P)X,\nonumber\\
 \dot{X}&=& (\alpha S+\beta P-1)X. 
\end{eqnarray}
where $\alpha$ and $\beta$ are the rate for a primary infection and the rate for
secondary infections, respectively. The recovery rate is set to one by suitable choice of time scale.

\begin{figure}
\centerline{\includegraphics[scale=.42]{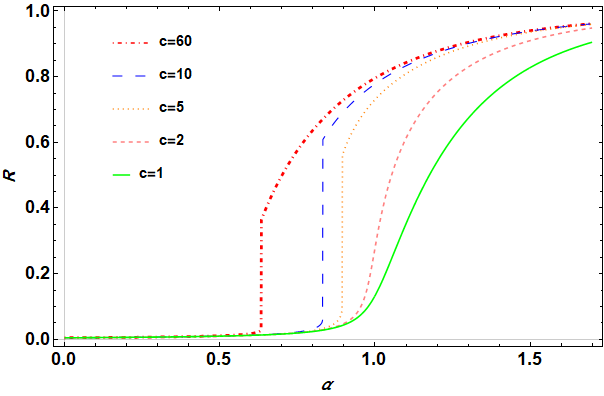}}
\caption{Phase transitions from disease-free to epidemic regime of two coupled SIR (CGCG model \cite{chen2013outbreaks}). Order parameter $R = 1 - S_{\infty}$ plotted against $\alpha$ for $\epsilon= 0.005$ . The curves correspond to different levels of cooperativity $c$.} 
\label{gcgc}
\end{figure}

In \cite{chen2013outbreaks} the authors have numerically studied the order parameter $R=1-S_\infty$ where $S_\infty= \lim_{t\rightarrow\infty}S(t)$ and have concluded that for some values of parameters $\alpha$ and $c$ one observes a discontinuity in the order parameter. They have sketched the order parameter as a function of $\alpha$ for different values of $c$ (see figure \ref{gcgc}).
For the initial conditions, they have considered 
\begin{eqnarray}
S(0)=1-\epsilon,\nonumber \\
P(0)=X(0)=\epsilon/2,
\end{eqnarray}
which means that only a very small fraction $\epsilon$ of the population have been infected. They have observed that smaller $\epsilon$'s lead to more clear discontinuities.

Our aim in this section is to solve the equations analytically and using the solution, derive (approximate) formula for the curves in figure \ref{gcgc} and also dependence of transition point on external parameters. Additionally we will shed insight on the nature of the transition and show that some non-trivial power-law relations exist in the system.

%%%%%%%%%%%%%%%%%%%%%%%%%%
\subsection{Exact solution of the equations}
%%%%%%%%%%%%%%%%%%%%%%%%%%

In all the equations \ref{main-eq}, the rate of changes is proportional to $X$, therefore we can introduce a new "time" variable defined via $d\tau=X(t)dt$. This new time variable has another interpretation as we will see later. Through this change of variable, the above equations turn out to be linear and can be solved exactly:
\begin{eqnarray}
\frac{dS}{d\tau}&=&-2 \alpha S,\nonumber \\
 \frac{dP}{d\tau}&=& \alpha S-\beta P,\nonumber\\
 \frac{dX}{d\tau}&=& \alpha S+\beta P-1.  
\end{eqnarray} 
The first equation is a very simple one and the solution is $S=S_0 \exp (-2\alpha\tau)$, where $S_0$ is given by the initial conditions. The second equation is also a linear equation in $P$ and as we have already obtained $S(\tau)$, the solution to this equation would be read as:
\begin{equation}
P(\tau)=P_0 e^{-\beta \tau} - \frac{\alpha S_0 }{\beta-2 \alpha} \left( e^{-\beta \tau}-e^{-2\alpha \tau}\right).
\end{equation}

\begin{figure}
\centerline{\includegraphics[scale=.3]{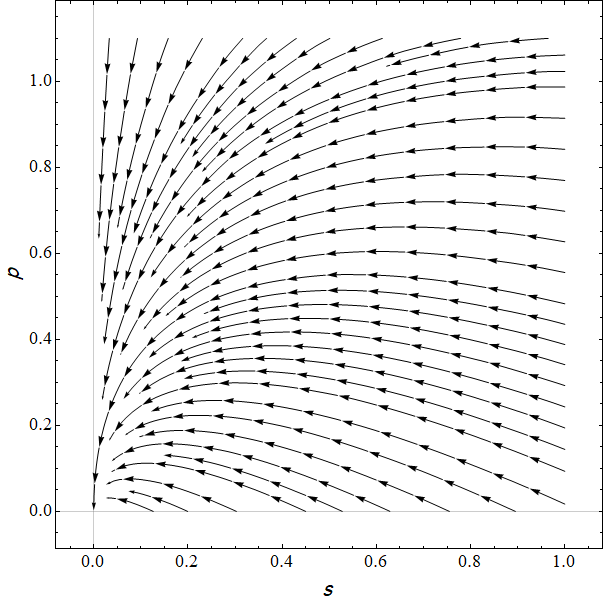}}

\centerline{\includegraphics[scale=.3]{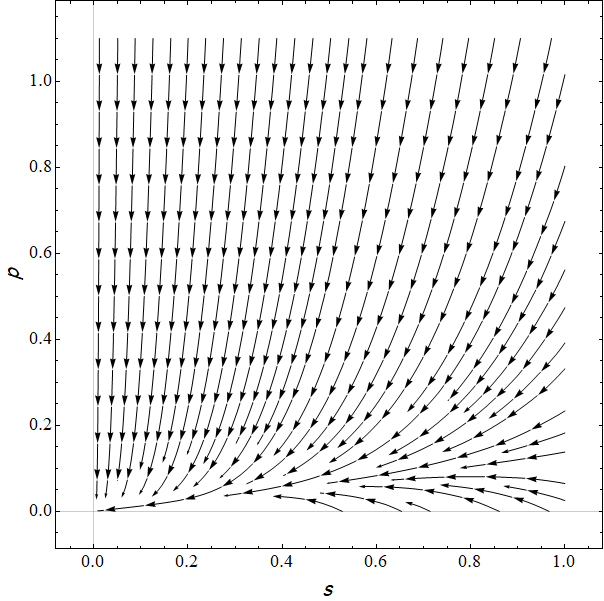}}
\caption{Phase portraits of the dynamical system in two-disease coinfection  for $\alpha=0.8$ and  $c=1$ (top) and $10$ (bottom). }
\label{Phase portraits}
\end{figure}

Before moving to solve the equation for $X$, we would like to have a better insight from the dynamics of these two variables. As the dynamics of $S$ and $P$ in terms of $\tau$ is independent of $X$ this can be done easily: We have a two-dimensional dynamical system and the best thing is to draw the phase portrait of the system in different cases. As in \cite{chen2013outbreaks}, we define the new parameter $c=\beta/\alpha$ which shows the level of cooperativity of the two diseases. Fig \ref{Phase portraits} shows the phase portrait of the dynamical system for $c=1,10$ and $\alpha=0.8$. From a simple analysis of eigenvectors and eigenvalues of the above (linear) dynamical system it can be observed that for $c<2$ this dynamical system approaches to origin in the direction (0,1) and for $c>2$ will approach the origin from the other eigenvalue $(c-2,1)$. Actually although our system follows the curves shown in the phase portraits, but do not necessarily end it at the origin. As we will explain bellow the path is terminated at some specific value of $\tau$.   
  
To have a better understanding  of the parameter $\tau$, we introduce the variable $U={\rm [a]+[aB]+[ab]=[b]+[Ab]+[ab]}$ which is the number of agents that have recovered from one of the diseases. It is straightforward to check that $S+P+X+U$ is the total number (fraction) of the agents and is conserved. Also we have $dU/dt=X$ or $dU= X(t) dt$. Comparing with the definition of our time parameter $\tau$, it is clear that the two parameter can be taken to be identical with suitable choice of initial values. During the dynamics, $U$ will rise from zero but stops to grow when the system reaches its final state. At this state surely $U\leq 1$ and hence one concludes that our new time variable, $\tau$, cannot continue to infinity, rather it will stop at a point where the total number of infective agents, $X$ vanishes. 

Let's turn to solve the equation governing $X$. As we already know $S+P+X+\tau=1$, one easily reads:
\begin{equation}\label{Xtau}
X(\tau)=1-\tau-S_0 e^{-2\alpha\tau}-P_0 e^{-\beta \tau} + \frac{\alpha S_0 }{\beta -2 \alpha } \left( e^{-\beta \tau}-e^{-2\alpha \tau}\right).
\end{equation} 
We assume that in the beginning there is no one who has already recovered from a disease so $\tau$ begins from zero. The next step is to find $\tau$ in terms of $t$ which in principle can be done through the integration $t=\int d\tau/X(\tau)$, but this integral could not be expressed in terms of known function. However, as we will see, the dependence of the variable on the actual time parameter $t$ plays little role.

%%%%%%%%%%%%%%%%%%%%%%%%%%
\subsection{Saddle-node bifurcation}
\begin{figure}
\centerline{\includegraphics[scale=.42]{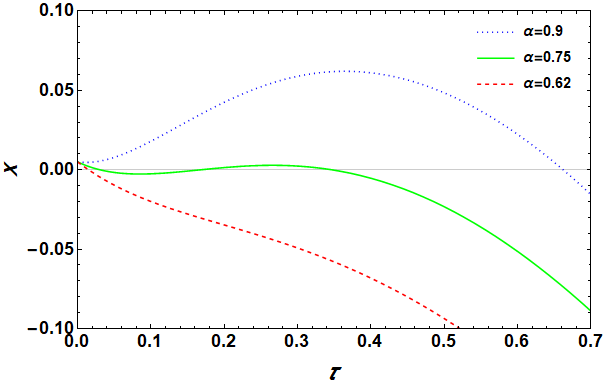}}
\caption{Fixed points, where $X(\tau_{\infty})=0$, of the dynamics. $X$ plotted against $\tau$ for $\epsilon= 0.01$ and $c=10$. The curve corresponds to different rates for a primary infection $\alpha$. Comparing the curves shows that there is a saddle-node bifurcation.}
\label{Xvstau}
\end{figure}

%%%%%%%%%%%%%%%%%%%%%%%%%%
Using the above exact solution, we try to obtain the same graphs. To find the order parameter, we have to find the value of $S$ at $t\rightarrow \infty$, or in terms of our time variable $\tau$ we have to find the value of $S$ at the point $\tau_\infty$ where the process stops. We call the point $\tau_\infty$ the fixed point. As stated, fixed point can be find via the relation $X(\tau_\infty)=0$ where $X(\tau)$ is given by Eq. \ref{Xtau}. As there are both exponential and polynomial terms in $X(\tau)$ the solution can not be found analytically, however we can understand what is the cause of the discontinuity: In figure \ref{Xvstau} the solution to $X(\tau)$ (Eq.\ref{Xtau}) is sketched for $c=10$, $\epsilon=0.01$ and for three different values of $\alpha$, 0.62, 0.75 and 0.90.   

With the specified initial conditions, the graph begins from $\epsilon/2$ and the initial slope is $\alpha-1+\epsilon\alpha(c-2)/2$ which is negative for $\alpha<1$ and sufficiently small values of $\epsilon$.  When $\alpha=0.62$, there is only one solution to $X(\tau)=0$. Clearly the solution is of the order of $\epsilon$ and therefore $\tau_\infty=O(\epsilon)$. This means that $R=1-S_\infty=1-S(\tau_\infty)$ is also of the order of $\epsilon$. As $\alpha$ is increased two other solutions appear ($\alpha=0.75$ curve in figure \ref{Xvstau}). But the initial condition of the dynamical system is $\tau=0$ and again we will end up in the smallest fixed point. The case of $\alpha=0.95$ is very different, there is again only one solution to the equation but the value of this solution is of the order of unity, therefore there will be a jump to a large value for $\tau_\infty$ when the smaller fixed points disappear. From the point of view of dynamical systems, a saddle-node bifurcation has happened. We can focus on the equation $\dot{\tau}=X(\tau)$ and ask if we begin from $\tau=0$ what will be the value of $\tau$ and $t\rightarrow\infty$. Clearly figure \ref{Xvstau} shows that there is a saddle-node bifurcation in the system as we change external parameters like $\alpha$ and hence a discontinuity is observed. It is well-known that when there is at least two stable fixed point in a dynamical fixed point, through a saddle-node type bifurcation there can happen a discontinuous transition \cite{Strogatz}. There are many examples of this type, a well-known of which is the outbreak of insects\cite{Ludwig1978}. In this system through changing the external parameters, the number of fixed point changes, first we have one stable fixed point which is related to a low population of insect. Changing the parameters, a saddle-node bifurcation happens and the system enters a bi-stable situation. However the system stays at the low-population fixed point until there is another saddle-node bifurcation through which this stable fixed point is vanished and the system suddenly jumps to the second (high-population) fixed point and a discontinuous transition is made. The situation in our problem is just the same but with one difference. The variable that jumps and stays at fixed point is a time variable and when it reaches the fixed point the dynamics is finished. Therefore it is not possible to observe phenomena like hysteresis  in the system. 

%%%%%%%%%%%%%%%%%%%%%%%%%%
\subsection{Transition points}
%%%%%%%%%%%%%%%%%%%%%%%%%%
At the bifurcation point that leads to the discontinuity, we should have 
\begin{equation}\label{Crit-Cond}
X(\tau)=\frac{dX}{d\tau}(\tau)=0
\end{equation}
 which cannot be solved analytically, though still a lot can be obtained at least in the limit where $\epsilon$ is very small, which is just the limit we are interested in. Let's call the value of $\tau$ where the bifurcation occurs by $\tau_{\rm crit}$. It is clear that in the limit $\epsilon\rightarrow 0$, $\tau_{\rm crit}$ becomes infinitesimal too. Therefore to obtain $\tau_{\rm crit}$ we can expand $X(\tau)$ in terms of $\tau$. To have a solution for equations \ref{Crit-Cond} we have to expand $X$ at least to second order of $\tau$. Keeping in mind that the expansion is in fact in terms of $\epsilon$ to the lowest order we find
\begin{eqnarray}\label{analyticcrit1}
\alpha_{\rm crit}(c,\epsilon)=1-\sqrt{\epsilon(c-2)}+O(\epsilon),\nonumber\\
\tau_{\rm crit}=\sqrt{\frac{\epsilon}{c-2}}+O(\epsilon).
\end{eqnarray}  

This result gives relatively accurate values for $\alpha_{\rm crit}$ and $\tau_{\rm crit}$, for example for $c=5$ and $\epsilon=0.001$ through numerical solving equations \ref{Crit-Cond} one obtains $\alpha_{\rm crit}\simeq 0.9485$ and $\tau_{\rm crit}\simeq 0.0207$ while our approximation yields $\alpha_{\rm crit}\simeq 0.9452$ and $\tau_{\rm crit}\simeq 0.0183$. It is straightforward to find the approximate solution up to order of $\epsilon$, we have to expand $X(\tau)$ up to cubic term in $\tau$ and keep all the terms which are of the order of $\epsilon^3$ which leads to 
\begin{eqnarray}\label{analyticcrit2}
\alpha_{\rm crit}(c,\epsilon)=1-\sqrt{\epsilon(c-2)}+\frac{2 c^2-5 c+4 }{3 (c-2)} \epsilon+O(\epsilon^{3/2}),\nonumber\\
\tau_{\rm crit}=\sqrt{\frac{\epsilon}{c-2}}+\frac{2 (2 c^2-5 c+4) }{3 (c-2)^2} \epsilon+O(\epsilon^{3/2}).
\end{eqnarray} 
For the above example $c=5$ and $\epsilon=0.001$ this gives $\alpha_{\rm crit}\simeq 0.9484$ and $\tau_{\rm crit}\simeq 0.0204$ which in the case of $\alpha_{\rm crit}$ is only $0.01$ percent off the correct answer. 
 
Note that for any $c>2$ we have $\lim_{\epsilon\rightarrow 0} \alpha_{\rm crit}=1$ and the transition only occurs at $\alpha=1$.  Also for any fixed value of $\epsilon$, as $c\rightarrow 2^+$ our expansion becomes useless due to the fact that higher order terms become more and more important in this limit. 

Also it worth to mention that there can be other symmetric initial conditions. For example, instead of having two distinct individuals who are infected by each of the diseases, we can consider that there is  just one agent carrying both diseases. In this case the initial conditions would be $S(0)=1-\epsilon$, $P(0)=0$ and $X(0)=\epsilon$. Following the above formalism, the transition point turns out to be $\alpha_{\rm crit}= 1-\sqrt{2\epsilon (c-2)}$ which is lower that the previous one. It is quite natural, because the infection rate is proportional to $X$ and in this case the initial value of $X$ is greater than the other case. 

%%%%%%%%%%%%%%%%%%%%%%%%%%
\subsection{Calculation of the order parameter: $R_\infty=1-S(\tau_\infty)$}
%%%%%%%%%%%%%%%%%%%%%%%%%%

Next we would like to find $R(\alpha)$ for $\alpha<\alpha_{\rm crit}$ in figure \ref{gcgc}. In this part $\tau_{\infty}<\tau_{\rm crit}$, and therefore is small. 
We expand $X(\tau)$ to second order of $\tau$ and solve the equation $X(\tau_\infty)=0$ which yields
\begin{equation}
\tau_\infty=\frac{\epsilon }{2(1-\alpha) }-\frac{\alpha(2-\alpha )   (c-2) \epsilon ^2}{8 (1-\alpha )^3}.
\end{equation} 
Then it is easy to find the order parameter for $\alpha<\alpha_{\rm crit}$: 
\begin{equation}
R=1-S(\tau_\infty)= (1-\epsilon)\exp(-2 \alpha \tau_\infty)
\end{equation}

If we find the order parameter for $\alpha>\alpha_{\rm crit}$ then the whole parts of figure \ref{gcgc} is at hand. In this case, $\tau_\infty$ is not small and the expansion does not work. However in some cases we can obtain some analytic results: suppose $c\gg 1$, then in the solution to $X(\tau)$ (equation \ref{Xtau}) we can neglect the terms proportional to $\exp{(-\beta \tau)}$. Also the parameter $\epsilon$ plays little role in this solution and can be dropped from all of our calculations. Through these simplifications one arrives at
\begin{equation}
X(\tau)=1-\tau-e^{-2\alpha \tau} \frac{c-1}{c-2}.
\end{equation}
Note that the factor $(c-1)/(c-2)$ can also be neglected for large values of $c$. We are looking for the solution of $X(\tau_\infty)=0$ which is of the order of unity to put it in $R=1-S(\tau_\infty)=1-\exp(-2 \alpha \tau_\infty)$, that is, neglecting the factor mentioned, we have $R=\tau$. This equation has a marvelous interpretation: when $c$ is large, in the end one can not find people who has been infected only by one of the diseases since 
\begin{eqnarray}
\tau=U&=&\left({\rm([ab]+[aB]+[a])+([ab]+[Ab]+[b])}\right)/2\nonumber\\
&=& {\rm [ab]+([aB]+[Ab])/2+([a]+[b])/2}
\end{eqnarray}
and $aB+Ab$ vanishes at $\tau_\infty$, this means that $a+b$ should vanish too. 

\begin{figure}\label{Compare}
\centerline{\includegraphics[scale=0.4]{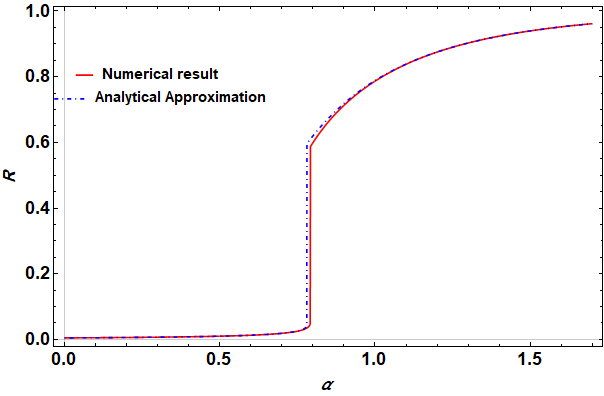}}
\caption{Comparison of the analytical approximation with the numerical result data for $c = 15$ and $\epsilon = 0.005$. We have intentionally took a relatively large value for $\epsilon$. For smaller values of $\epsilon$ the difference of the two curves was not observable. }

\end{figure}

To obtain an approximate solution, we write $\tau_\infty=1-\delta$ and solve
$X(\tau_\infty)=0$ up to quadratic terms of $\delta$

\begin{equation}
R=\frac{-\sqrt{-4 \alpha ^2-4 e^{2 \alpha } \alpha +e^{4 \alpha }}-2 \alpha +e^{2 \alpha }}{4 \alpha ^2}.
\end{equation}

 and the second branch of the order parameter diagram is obtained. In \cite{chen2013outbreaks} the authors had observed that this branch is not so dependent on $c$ for large values of this parameter; this is what we have already observed, for $c\gg 1 $ the parameter $\beta$ can be neglected and the top branches coincide for different $c$'s. Figure \ref{Compare} puts all the derived results in comparison with the numerical data for $c=15$ and $\epsilon=0.005$. The difference is so little especially when we are far from transition point, actually if we take  $\epsilon=0.0001$ one could not distinguish the two curves within such a graph.
  
%%%%%%%%%%%%%%%%%%%%%%%%%%
\subsection{Hybrid transitions}
%%%%%%%%%%%%%%%%%%%%%%%%%%
The next point is about the hybrid nature of the transition. Although the order parameter changes discontinuously at the transition it can be argued that just below the transition point a power-law behavior is observed. The transition point happens at the saddle-node bifurcation point. Near this point the function $X(\tau)$ can be approximately written as $X(\tau)= a(\alpha)+ b(\tau-\tau_{\rm crit})^2$ where $a(\alpha)=a_0 (\alpha-\alpha_{\rm crit})$ and $a_0$ is a constant. Therefore when $\alpha$ tends to $\alpha_{\rm crit}$ from below, one can read $\tau_\infty$ as
\begin{equation}
\tau_\infty- \tau_{\rm crit}=\sqrt{\frac{a_0}{b}}(\alpha_{\rm crit}-\alpha)^{(1/2)} 
\end{equation}
As the order parameter $R$ is just $1-S(\tau_\infty)$,  it will be easy to conclude  $R(\alpha_{\rm crit})-R(\alpha)\propto \sqrt{\alpha_{\rm crit}-\alpha}$. This nontrivial power-law can help us as an alarm of approaching the transition: $dR/d\alpha$ diverges as we tend to the transition point.

%%%%%%%%%%%%%%%%%%%%%%%%%%
\subsection{Another approach: minimums of the potential}
\begin{figure}
\centerline{\includegraphics[scale=0.40]{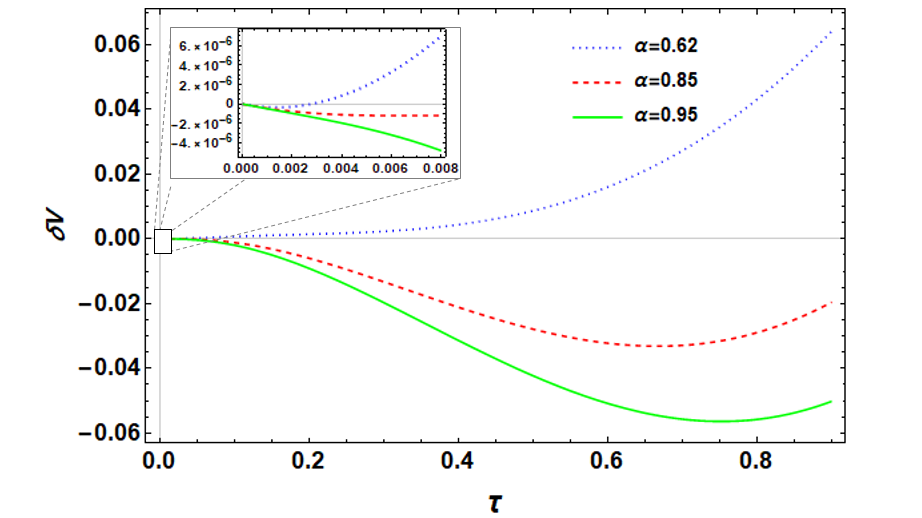}}
\caption{Potential $\delta V$ plotted against $\tau$ for $\epsilon= 0.001$ and $c=30$ . The curves correspond to different rates for the primary infection $\alpha=0.62,\,0.85,\, 0.95$. The inset graph shows a larger view for very small values of $\tau$. It is clear that for $\alpha=0.62$ there is a minimum near $\tau=0$ which vanishes for larger $\alpha$'s.}
\label{Potential}
\end{figure}

%%%%%%%%%%%%%%%%%%%%%%%%%%
In the end of this section we would like to refer that some parts of this exact results have been obtained in a different theme before. In \cite{janssen2016first} they have considered a similar problem and have integrated out the two first equations as we have done. Then using the equation $d\tau/dt=X(\tau)$ and having the solution of $X(\tau)$ they have re-expressed this equation in the following way:
\begin{equation}
\frac{d\tau}{dt}=X(\tau)=-\frac{dV(\tau)}{d\tau},
\end{equation}
 that is, the dynamics is treated as the dynamics of a particle in a viscous media moving in the presence of the potential $V(\tau)$. Because of the dynamics, the particle always finds the local minimum  and if the minimum is vanished by changing external parameter, it will go to the next local minimum which leads to a discontinuous transition. This potential is shown for three different values of $\alpha$, with $\epsilon=0.001$ and $c=30$ in figure \ref{Potential}, where by increasing $\alpha$ the first local minimum is vanished. In \cite{janssen2016first} they have find very similar results to ours, although from their point of view and the questions they have answered are different.

%%%%%%%%%%%%%%%%%%%%%%%%%%%%%%%%%%%%%%%%%%%%%%%%%%%%%%%%%%%%%%%%%%%%%%%%%%%%%%%%%%%%%%
\section{Generalization to three or more diseases}
%%%%%%%%%%%%%%%%%%%%%%%%%%%%%%%%%%%%%%%%%%%%%%%%%%%%%%%%%%%%%%%%%%%%%%%%%%%%%%%%%%%%%%

\begin{figure}
\centerline{\includegraphics[scale=.6]{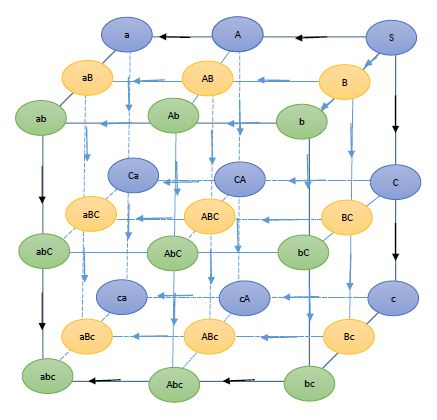}}
\caption{Schematic of three-disease coinfection with A,B,C symmetry and restrictions on the infection rates as discussed in the text. Capital letters A, B and C represent infective states,
lower-case letters a,b and c stand for “recovered” ones.}
\label{3dis}
\end{figure}

In this section we will generalize the problem to three or more diseases. For simplicity we begin with the case where the number of diseases, $n$ is 3, and then the generalization to arbitrary $n$ is brought. Figure \ref{3dis} shows schematically the agents considered at the model and how they are transformed into one another. Again capital letters show the agents that are infected with the disease and small letter shows that the agent has already become immune to the disease. For example [AbC] means that the agent is infected with diseases A and C and has already become immune to b.

Again,  Assuming the diseases have identical properties, the equations governing the dynamics would be symmetric and one can introduce the following new variables: $P_0=[S]$, the fraction of agents who have not infected with any of diseases, $P_1{\rm =[a]+[A]=[b]+[B]=[c]+[C]}$ the fraction who have or have had only one of the diseases, $P_2={\rm [aB]+[Ab]+[ab]+[AB]}=\ldots$ the fraction of agents who have or have had exactly two diseases and
 $X={\rm[A]+[Ab]+[AB]+[Ac]+[AC]+}$ ${\rm [Abc]+[ABc]+[AbC]+[ABC]}=\ldots$
  the fraction of agents can transfer the diseases A (or equivalently B or C). We call these groups susceptible, 1-disease, 2-disease and infective group respectively. The primary, secondary and tertiary infection rates are taken to be $\beta_0$, $\beta_1$ and $\beta_2$ and independent of the type of disease the agent has been infected with. That means, any individual who has not experienced any diseases, in presence of an infective person,  will be infected the first disease with rate $\beta_0$, any agent who has experienced (and possibly has become immune to) one disease will be infected by a second disease with rate $\beta_1$ and those who have already been infected with two diseases will get the third one with rate $\beta_2$. Therefore, the rate of changes of $P_0$ is simply proportional to $\beta_0 X P_0$, but there are three ways to be infected we have $\dot{P}_0=-3 \beta_0 P_0 X$. Also it is clear that the rate of changes in $P_1$ has two terms, one proportional to $P_0 X$ and the second proportional to $P_1 X$. The former is the number of agents who have already experienced one disease and are infected with a second one, and the latter is the number of agents that have experienced two diseases and now get the third one. With similar reasoning one arrives at the following equations 
\begin{eqnarray}
\label{3eq1}
\frac{dP_0}{dt}&=&-3\beta_0 P_0 X\nonumber\\
\frac{dP_1}{dt}&=&(\beta_0 P_0 -2\beta_1 P_1) X\nonumber\\
\frac{dP_2}{dt}&=&(2\beta_1 P_1 -\beta_2 P_2) X\nonumber\\
\frac{dX}{dt}&=&(\beta_0 P_0 +2\beta_1 P_1+\beta_2 P_2-1) X.
\end{eqnarray}    
The numerical pre-factors can be computed easily: There are three ways that a susceptible person is infected, so we have a factor 3 in the first equation. There is only one way to arrive at a specific 1-disease group from susceptible group and two ways to go out from 1-disease group to a 2-disease group, and hence the numerical factors of the second equation is obtained. For the third equation a similar argument can be made. The dynamics of $X$ can be obtained in the following way: In figure \ref{3dis} consider the horizontal plane which has capital C in all of the vertices, which represents $X$. The flow toward this plane, which gives the positive terms, comes from the upper plane which has one $P_0$, two $P_1$'s and one $P_2$. The flow outward the plane is just $-X$ as we have fixed the recovery rate to be unity by adjusting time scale.  From equation of \ref{3eq1} one can obtain 
\begin{equation}
\frac{d(P_0+2P_1+P_2+X)}{dt}=-X.
\end{equation}

Again it is observed that the rates of changes in all the variables is proportional to $X$ and hence a new time variable can be defined via $Xdt=d\tau$ to make all the equations linear. It turns out that the result is qualitatively the same as the case of two diseases. There will be a discontinuous transition as we increase $\alpha=\beta_0$ while keeping $c_1=\beta_1/\beta_0$ and $c_2=\beta_2/\beta_0$ fixed. Following  the steps explained in section \ref{analytic} the transition point is obtained to be $\alpha_{\rm crit}=1-\sqrt{\epsilon((4/3) c_1 -2)}$ which is independent of $c_2$. This turns out to be a general feature as we will see below. Also note that the transition may occur when $c_1>3/2$ where in the 2-diseases situation only for $c>2$ the transition happened in the system.

The generalization to $n$ diseases is straightforward. When system is symmetric, one introduces the variables $P_m$ which are the fraction of agents having experienced exactly $m$ diseases ($0\leq m<n$) and the variable $X$ which is the fraction of agents transferring one specific disease. As in each time step, an agent either is infected with a new disease or recovered from one of the diseases the dynamics will be
\begin{equation}
\frac{d P_m}{dt}= \left(m \beta_{m-1}P_{m-1}-(n-m)\beta_m P_m\right)X, 
\end{equation}        
where $\beta_m$ is the infection rate of the ($m+1$)th disease when the agent has already experienced $m$ diseases. The numerical factors can be obtained easily by considering how many ways are there to go from a specific ($m-1$)-disease group to a $m$-disease group. Also for the variable $X$ we have
\begin{equation}
\frac{dX}{dt}= \left(-1+\sum_{m=0}^{n-1} 
\left(
\begin{array}{c}
 n-1 \\
 m \\
\end{array}
\right)\beta_{m} P_m\right)X
\end{equation}
As before through introducing $d\tau=Xdt$ the above equations will become solvable and the transition point can be obtained in terms of $\epsilon$ and the $c$ parameters. For example for $n=4$ we arrive at 
\begin{equation}
\alpha_{\rm crit}=1- \sqrt{\epsilon\frac{3c_1-4}{2}}.
\end{equation} 
where the lowest value of $c_1$ that the transition is present in the system is $4/3$ which is lower than the cases $n=2,3$. If we adopt the conjecture that $\alpha_{\rm crit}$ is independent of $c_2, c_3,\ldots$ for arbitrary $n$, we may set all these parameters to zero and compute $\alpha_{\rm crit,n}$

\begin{eqnarray}
\alpha_{\rm crit,n}=1-\sqrt{\frac{2\epsilon}{n}\left(c(n-1)-n\right)},
\end{eqnarray}    
which reveals the minimum $c$ to have a discontinuity to be $n/(n-1)$.

%%%%%%%%%%%%%%%%%%%%%%%%%%%%%%%%%%%%%%%%%%%%%%%%%%%%%%%%%%%%%%%%%%%%%%%%%%%%%%%%%%%%%%
\section{Asymmetry Considerations}
%%%%%%%%%%%%%%%%%%%%%%%%%%%%%%%%%%%%%%%%%%%%%%%%%%%%%%%%%%%%%%%%%%%%%%%%%%%%%%%%%%%%%%

So far, we have considered completely symmetric systems, however such systems can only be approximations to the real world systems. Therefore it is very important to see if breaking symmetry will affect the general features of the model or not. We will introduce asymmetry to the model in two ways: first we suppose the dynamic is symmetric while the initial conditions is not. Second we will change the parameters of different diseases so that the dynamics be asymmetric. 

%%%%%%%%%%%%%%%%%%%%%%%%%%
\subsection{Asymmetry of the initial conditions}
%%%%%%%%%%%%%%%%%%%%%%%%%%

Let's for simplicity consider the case of two diseases. If at the beginning the number of agents infected by disease A is different from the number of agents infected by disease B, even though the dynamics is symmetric, the variables show no symmetry and the simplifications we have taken into account does not work any more. In particular we have to introduce two distinct $P$'s and two distinct $X$'s for the system:
\begin{eqnarray}
P_{A}&=&{\rm [A]+[a]},\nonumber\\
P_{B}&=&{\rm [B]+[b]},\nonumber\\
X_{A}&=&{\rm [A]+[Ab]+[AB]},\nonumber\\
X_{B}&=&{\rm [B]+[aB]+[AB]}.
\end{eqnarray}  
When $X_A$ and $X_B$ are different, we are not able to introduce our new time scale consistently to make the equations linear. However the question we would like to answer is that if the initial values of the equations are a bit asymmetric, does this asymmetry grow with time or it will fade away. To begin we write the equations of motion in asymmetric case
\begin{eqnarray}
\label{asymdyn}
\frac{dS}{dt} &=& -( \alpha_A X_A +\alpha_B X_B )S,  \nonumber \\
\frac{dP_A}{dt} &=& \alpha_A X_A S -\beta_B X_B P_A, \nonumber \\
\frac{dP_B}{dt} &=&  \alpha_B X_B S -\beta_A X_A P_B, \nonumber \\
\frac{dX_A}{dt} &=&  \alpha_A X_A S +\beta_A X_A P_B -X_A,  \nonumber \\
\frac{dX_B}{dt} &=&  \alpha_B X_B S +\beta_B X_B P_A -X_B.  
\end{eqnarray}
in the symmetric case we have the solution $X_A(t)=X_B(t)=X(t)$ and $P_A(t)=P_B(t)=P(t)$. We assume $X_{A/B}(t)=X(t)+\epsilon_{A/B}(t)$ and $P_{A/B}(t)=P(t)+\delta_{A/B}(t)$ and put them in equations (\ref{asymdyn}) and expand in terms of $\epsilon_{A/B}$ and $\delta_{A/B}$ and keep the linear terms in these functions. The result can be written in the matrix form:
\small
\begin{eqnarray}\label{matrix}
\hspace{-0.5cm} \left(
\begin{array}{cc}
\dot{\delta}_A\\\dot{\delta}_B\\\dot{\epsilon}_A\\\dot{\epsilon}_B\\
\end{array}
\right)
&=&\nonumber\\
&&\hspace{-2.0cm}
\left(
\begin{array}{cccccc}
 -\beta X & 0 & \alpha S & -\beta P \\
 0 & -\beta X  & -\beta P & \alpha S \\
 0 & \beta X  & \alpha S+\beta P-1  & 0 \\
 \beta X  & 0 & 0 & \alpha S+\beta P-1 \\
\end{array}
\right)
\left(
\begin{array}{cc}
\delta_A\\ \delta_B\\ \epsilon_A\\ \epsilon_B\\
\end{array}
\right)
\end{eqnarray}
\normalsize
or in a compact form
\begin{equation}
\frac{d\vec{L}}{dt}=G \vec{L}. 
\end{equation}
If the largest real part of the eigenvalues of the matrix $G$ is negative then the symmetric solution is an attractive fixed state. Note that the elements of the above matrix are functions of time and so are the eigenvalues. Hopefully the eigenvalues can be obtained in terms of the functions $S$, $P$ and $S$; 
\small
\begin{eqnarray}
\lambda_{1,2}&=&\frac{1}{2} \left(\alpha  S+\beta P-\beta X-1\right.\nonumber\\&& \hspace{-1cm} \left. \pm\sqrt{(-\beta P+\beta X-\alpha  S+1)^2-4 \beta X}\right),\nonumber\\
\lambda_{3,4}&=&\frac{1}{2} \left(\alpha  S+\beta P-\beta X-1\right.\nonumber\\
&&\hspace{-1cm}\left. \pm\sqrt{(-\beta P+\beta X-\alpha  S+1)^2-4 \left(\beta X-2 \alpha \beta S X\right)}\right).
\end{eqnarray}
\normalsize

Therefore  using the solution obtained for these functions in the section II, the eigenvalues and eigenvectors are at hand in terms of our parameter $\tau$, the expression for arbitrary parameters is very long and we do not bring it here. In fact, it is enough to note some general properties of the eigenvalues and eigenvectors of $G$:  Two of the eigenvectors associated with eigenvalues $\lambda_{1,2}$ are in the form $(g,-g,1,-1)$ with $g$ being a function of $S$, $P$,  and $X$ while the other two eigenvectors are in the form $(h,h,1,1)$, again with $h$ a function of $S$, $P$ and $X$. In other words there are two subspaces that are invariant under the effect of the matrix introduced in eq. \ref{matrix}. For initial values we consider that the number of agents infected with disease A is a bit higher (lower) than those infected with disease B. Therefore the system evolves only in the subspace produced by the vectors $(g,-g,1,-1)$ and the other two eigenvalues could be neglected. 

The eigenvalues $\lambda_{1,2}$ are the solution to the quadratic equation $\lambda^2-W \lambda +\beta X=0$ with $W=\alpha S+\beta P -\beta X -1$. Therefore when $X=0$, one of the eigenvalues becomes zero. In fact when $X=0$ the matrix $G$ is degenerate with two eigenvalues equal to zero and the other two equal to $\alpha S+\beta P -1=W|_{X=0}$. The parameter $W$ is negative in the beginning of the dynamics, it equals to $\alpha-1-\epsilon$. Therefore at the beginning of the dynamics, surly the symmetric solution is attractive.   

\begin{figure}
\centerline{\includegraphics[scale=.42]{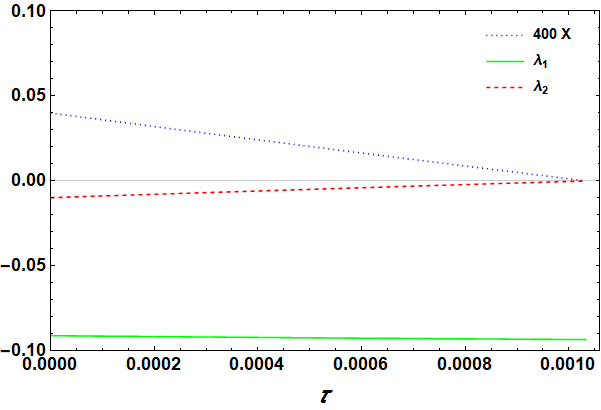}}
\vspace{5mm}
\centerline{\includegraphics[scale=.42]{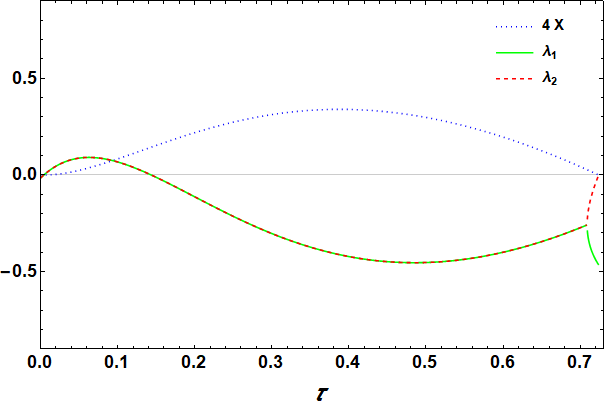}}
\caption{Asymmetric initial conditions. Real parts of the eigenvalues of the matrix $G$ for two cases, which shows if the symmetric solution is an attractive fixed point: (Top) below transition point with $c=15$, $\epsilon=10^{-4}$ and $\alpha=0.9$. (Bottom) above transition point with the same values for $c$ and $\epsilon$ but with $\alpha=0.97$.
In both graphs $X$ is also sketched. For the top graph, the real part o the eigenvalues are negative for $\tau<\tau_{\infty}$. For the case above the transition point, at the beginning the real part of the eigenvalues are positive however in most parts of the dynamics they are negative. Note that $X$ is plotted 400 (top) and 4 (bottom) times larger, in order to be seen.}
\label{eigen}
\end{figure}

 In fig. \ref{eigen} the two relevant eigenvalues are plotted against $\tau$ for two cases: (a) below the transition point with $c=15$, $\epsilon=10^{-4}$ and $\alpha=0.90$; (b) above the transition point with $c=15$, $\epsilon=10^{-4}$ and  $\alpha=0.97$. In these graphs $X$ is also plotted to explicitly see when $X$ vanishes and the system reaches its final state. For the cases below the transition point, the real part of these eigenvalues are negative for all $\tau<\tau_{\infty}$. Therefore the system is stable in this region.

The treatment of the cases that the diseases spread throughout the system is more tricky.  The real part of the eigenvalues become positive for an interval of $\tau$, but again it becomes negative for the rest of dynamics. Note that for most of $\tau$'s the real part of the two eigenvalues are equal. This means that for the distance from the symmetric solution changes isotropically within the subspace most of the time. To find how much the system is deviated from the symmetric solution, one should integrate  the changes of the variables over time, that is $\delta L_{tot}= \exp(\int dt\Re(\lambda_{1,2}))L_0$ or equivalently  $\delta L_{tot}= \exp(\int d\tau\Re(\lambda_{1,2})/X)L_0$. Therefore in fig. \ref{eigen} when $X$ is smaller the effect of $\lambda$ is greater; that is the beginning and the end of the dynamics, where $X$ is very small, are the most important parts. As stated before, at the beginning the real part of the eigenvalues are negative. At $\tau\rightarrow \tau_{\infty}$ where again $X$ is extremely small, the two eigenvalues tend to $0^-$ and $\alpha S(\tau_{\infty})+\beta P(\tau_{\infty}) -1\simeq 2\alpha\exp (-2\alpha \tau_{\infty})-1$ which is strictly negative. Therefore the symmetric solution is stable for fluctuations that break its symmetry. 

%%%%%%%%%%%%%%%%%%%%%%%%%%
\subsection{Asymmetry of the parameters}
%%%%%%%%%%%%%%%%%%%%%%%%%%
Another way to break symmetry is to change the parameters of the system. Thus we consider infection rates of two diseases are different. Then we break the symmetry in the following form: we suppose that one of the diseases spreads $k$ times more easily, that is the infection rates of the disease A and B are related to each other via $(\alpha_A,\beta_A)=k(\alpha_B,\beta_B)$. It is clear that $k=1$ corresponds to the symmetric case, and if $k\gg 1$  the first disease dominates and we can neglect the second disease. That is for $k=1$ a discontinuous transition may happen while for large 
values of $k$ only continuous transition can be found. It is interesting to investigate the intermediate steps to see how discontinuity vanishes as we increase $k$.

\begin{figure}
\centerline{\includegraphics[scale=.42]{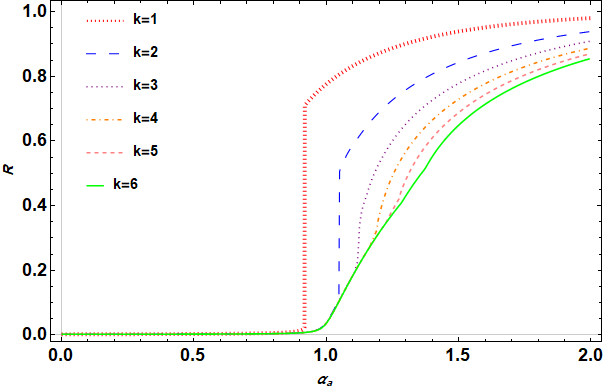}}
\caption{The order parameter of asymmetric system as a function of $\alpha_B$ (the smaller infection rate) for $k=1,2,\ldots, 6$ where $(\alpha_A,\beta_A)=k(\alpha_B,\beta_B)$ and $c=10$, $\epsilon=10^{-3}$. The discontinuity becomes smaller as $k$ increased and vanishes at $k\simeq 4$. }
\label{unsym}
\end{figure}

Fig. \ref{unsym} shows the order parameter as a function of $\alpha_A$ (which is the greater infection rate) for such systems with $c=10$, $\epsilon=10^{-3}$ and different $k$'s. The discontinuity is present in the model for $k\lesssim 4$ but the amount of jump becomes smaller as $k$ approaches $4$. Note that the place of discontinuity in terms of $\alpha_A$ has become more than what we had in symmetric case while $\alpha_B$ turns out to be less than the case of symmetric situation. In terms of $\alpha_A$ the transition point has become more than unity, the value where the continuous transition happen. The line of this continuous transition is visible in the graph. This means that the first disease undergoes the continuous transition first, then another transition, which is discontinuous happens at larger $\alpha$.   

\begin{figure}[b]
\centerline{\includegraphics[scale=.42]{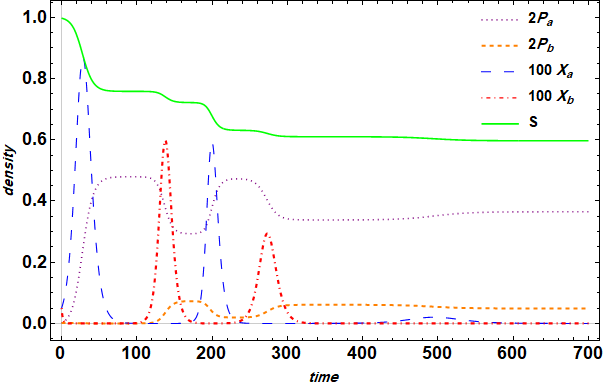}}
\caption{Time evolution of the variables $S$, $P_A$, $P_B$, $100\times X_A$ and $100\times X_B$ as functions of time for a system with $\epsilon=0.001$, $\alpha_B=0.38$, $c=10$ and $k=3$. Note that $\alpha_A=k\times \alpha_B=1.11>1$ . There are five different and disjoint peaks for infecting populations.  }
\label{2dis-asym}
\end{figure}

Let's also see how the variables evolve through time. It is interesting because now we have two distinct diseases that spread with different time scales and therefore the variables $X_A$ and $X_B$ would peak at different times. In general we have two distinct peeks in the graphs of $X_A$ and $X_B$ where expectedly the peak of $X_A$ comes first. However near the transition point very interesting phenomenon happens.   Fig. \ref{2dis-asym} shows the evolution of different variables for such a system with $k=3$, $c=10$ and $\alpha=0.38$. It is observed that there exist three distinct peaks for  $X_A$ and two for $X_B$. Note that $X_A$ and $X_B$ has been multiplied with a factor of 100 for a better visualization. Therefore $S$ decreases through five steps, each step is associated with one of the peaks. 

This can be understood in the following way. As $\alpha_A$ is larger than one, it will rise even if $B$ is not present in the system. When the number of agents experiencing the disease $A$ rises, there are a considerable amount of individuals susceptible for the second disease with a much higher rate $\beta_B=c \alpha_B$. This gives rise to number of agents infected with $B$ and the second peak appears. Now again there are a lot of agents already infected with $B$ and ready to be infected with $A$ as a second disease with a high rate  $\beta_A=c \alpha_A$ and causing a second peak for $X_A$. This may happen again and in this way several peaks may happen. The same phenomenon, and in fact a more complicated one, happens when there are more than two diseases are present in the theory. 

%%%%%%%%%%%%%%%%%%%%%%%%%%%%%%%%%%%%%%%%%%%%%%%%%%%%%%%%%%%%%%%%%%%%%%%%%%%%%%%%%%%%%%
\section{Summary and Discussion}
%%%%%%%%%%%%%%%%%%%%%%%%%%%%%%%%%%%%%%%%%%%%%%%%%%%%%%%%%%%%%%%%%%%%%%%%%%%%%%%%%%%%%%
In summary, we introduced and investigated a generalized framework for co-infection. We have studied the exact analytically solutions of cooperative coupled SIR dynamics for two and more pathogens (or strains) in mean-filed approximations. We calculated the epidemic threshold, i.e. transition point ($\alpha_{crit}$), the percentage of final infected host population, i.e. order parameter ($R_{\infty}$), for symmetric and asymmetric scenarios, and also one critical exponent. Moreover, we showed: 1) There is a saddle-node bifurcation point when two pathogens co-infect. 2) The discontinuous transitions disappear when asymmetry, in parameter space, between two pathogens become large enough. 3) This symmetry breaking also can lead to a greater epidemic threshold and smaller $R_{\infty}$. 3) Breaking symmetry of initial conditions, the symmetric solution is attractive at the beginning of the dynamics and it's stable for the fluctuations breaking the symmetry when $\tau\rightarrow \tau_{\infty}$. 4) For the case of $n$ cooperative diseases, the transition point is independent of the second ($c_2$), third ($c_3$), ... cooperation ratios. It also reveals the minimum $c$ (the first cooperation ratio) to have a discontinuity at $\frac{n}{n-1}$. 

We can conclude several points from our analysis in comparison with super-infection, see chapter 11 in \cite{nowak2006evolutionary}. Super-infection means that an already infected host can be infected by another pathogen which replaces the primary one(s). In contrast, co-infection means that an already infected host can be easier infected by other pathogens while they all co-exist in the host body. In other words, in co-infection scenario, the primary infections facilitate other infections. And while super-infection triggers intra-host competition, co-infection triggers intra-host cooperation. While super-infection increases the average level of virulence for pathogens, co-infection can decreases the epidemic threshold and increases the average level of pathogens' populations in comparison to non-interacting spreading dynamics. In super-infection scenario even the pathogen with highest reproductive ratio can extinct. In host population, all the pathogens can coexist in co-infection case with any infection and recovery rates, also in super-infection, pathogens with different level of virulence can coexist. A high virulent pathogen, which could not persist alone, can survive in super-infection and a very low transmissible pathogen, which could not make any outbreak alone, can cause an epidemic in host population when co-infects. Super and co- infections lead to dramatic change in the average level of affected host populations in opposite directions. The higher the rate of super-infection, the smaller the number of infected hosts; while the higher the rate of co-infection, the larger the number of infected hosts; Nevertheless there is an upper bound for occupation of the host population for given set of parameters as calculated.

We have limited our analysis to only cooperative SIR dynamics. We expect even richer dynamics from mixing different super- and co-infection dynamics, and believe these results could help to understand more complex scenarios of disease ecology \cite{hudson2002ecology,schmidt2001biodiversity,ostfeld2010infectious}. Also this analytical approach can open a road to or help to generalize the works which study mechanisms leading to discontinuous phase transitions at threshold such as different percolations \cite{achlioptas2009d,bizhani2012discontinuous,goltsev2006k,janssen2004generalized}, the cooperative complex contagion \cite{dodds2004universal} and cascades on interdependent networks \cite{buldyrev2010catastrophic,parshani2010interdependent,son2012percolation}.
 
%%%%%%%%%%%%%%%%%%%%%%%%%%%%%%%%%%%%%%%%%%%%%%%%%%%%%%%%%%%%%%%%%%%%%%%%%%%%%%%%%%%%%%
\section{Acknowledgment}
%%%%%%%%%%%%%%%%%%%%%%%%%%%%%%%%%%%%%%%%%%%%%%%%%%%%%%%%%%%%%%%%%%%%%%%%%%%%%%%%%%%%%%  
FGh acknowledges support by German Research Foundation (DFG) under grant GH 176/1-1, within the idonate program (project 345463468).

%%%%%%%%%%%%%%%%%%%%%%%%%%%%%%%%%%%%%%%%%%%%%%%%%%%%%%%%%%%%%%%%%%%%%%%%%%%%%%%%%%%%%%
\section*{References}
%%%%%%%%%%%%%%%%%%%%%%%%%%%%%%%%%%%%%%%%%%%%%%%%%%%%%%%%%%%%%%%%%%%%%%%%%%%%%%%%%%%%%%
\bibliography{paper.bib}

\end{document}